\documentclass[preprint]{aastex_mst}
\usepackage{natbib}
\usepackage{mathrsfs}
\usepackage{rotating}
\usepackage{graphicx}
\usepackage{longtable,lscape}

\newcommand{\etal}{et~al.}

\newcommand{\fn}[1]{\footnote{\scriptsize{#1}}}
\newcommand{\Eqn}[1]{Eq{#1}.}  
\newcommand{\Fig}[1]{Fig{#1}.}  
\newcommand{\Cassit}{\textit{Cassini}}  

\slugcomment{Submitted to \textit{The Astronimcal Journal}, 16 November 2009}

\shorttitle{}
\shortauthors{Tiscareno \etal}

\begin{document}

\title{\vspace{-0.15in}An analytic parameterization of\\self-gravity wakes in Saturn's rings,\\with application to occultations and propellers}
\author{Matthew~S.~Tiscareno,$^1$ Randall~P.~Perrine,$^2$ Derek~C.~Richardson,$^2$ Matthew~M.~Hedman,$^1$ John~W.~Weiss,$^{3,4}$ Carolyn~C.~Porco,$^3$ and Joseph~A.~Burns$^{1,5}$}
\affil{$^1$Department of Astronomy, Cornell University, Ithaca, NY 14853\\$^2$Department of Astronomy, University of Maryland, College Park, MD 20742\\$^3$CICLOPS, Space Science Institute, Boulder, CO 80301\\$^4$Physics and Astronomy Department, Carleton College, Northfield, MN 55057\\$^5$Department of Theoretical and Applied Mechanics, Cornell University, Ithaca, NY 14853}

\begin{abstract}

We have developed a semi-analytic method of parameterizing $N$-body simulations of self-gravity wakes in Saturn's rings, describing their photometric properties by means of only six numbers:  three optical depths and three weighting factors.  These numbers are obtained by fitting a sum of three gaussians to the results of a density-estimation procedure that finds the frequencies of various values of local density within a simulated ring patch.  

Application of our parameterization to a suite of $N$-body simulations implies that rings dominated by self-gravity wakes appear to be mostly empty space, with more than half of their surface area taken up by local optical depths around~0.01.  Such regions will be photometrically inactive for all viewing geometries.  While this result might be affected by our use of identically-sized particles, we believe the general result that the distribution of local optical depths is trimodal, rather than bimodal as previous authors have assumed, is robust. 

The implications of this result for the analysis of occultation data are more conceptual than practical, as we find that occultations can only distinguish between bimodal and trimodal models at very low opening angles.  Thus, the only adjustment needed in existing analyses of occultation data is that the model parameter $\tau_{gap}$ should be interpreted as representing the area-weighted average optical depth within the gaps (or inter-wake regions), keeping in mind the possibility that the optical depth within those inter-wake regions may vary significantly. 

The most significant consequence of our results applies to the question of why ``propeller'' structures observed in the mid-A~ring are seen as relative-bright features, even though the most prominent features of simulated propellers are regions of relatively low density.  Our parameterization of self-gravity wakes lends preliminary quantitative support to the hypothesis that propellers would be bright if they involve a local and temporary disruption of self-gravity wakes.  Even though the overall local density is lower within the propeller-shaped structure surrounding an embedded central moonlet, disruption of the wakes would flood these same regions with more ``photometrically active'' material (i.e., material that can contribute to the rings' local optical depth), raising their apparent brightnesses in agreement with observations.  We find for a wide range of input parameters that this mechanism indeed can plausibly make propellers brighter than the wake-dominated background, though it is also possible for propellers to blend in with the background or even to remain dark.  We suggest that this mechanism be tested by future detailed numerical models. 

\end{abstract}

\textit{Subject headings:}  Saturn, rings \\
\indent{}\textit{Running header:}  Analytic parameterization of self-gravity wakes

\section{Introduction \label{Intro}}
Saturn's dense A~and B~rings are pervaded by a microstructure, dubbed ``self-gravity wakes,'' of alternating dense and rarified regions that arise due to a rough balance between the clumping together of particles under their mutual self-gravity and their shearing apart again due to tidal forces \citep{JT66,Salo92,Richardson94}.  The characteristic alignment of elongated self-gravity wake structures causes variation with viewing geometry in the brightness of the rings as seen in ground-based optical \citep{Camichel58,Salo04,French07} and microwave observations \citep{Dunn02,Dunn04} as well as spacecraft images \citep{Franklin87,Porco08}, and in the rings' opacity as mesaured by stellar occultations \citep{Colwell06,Colwell07,Hedman07}, which in the latter case has led to empirical determination of some wake properties.  

Existing analyses of stellar occultation data from \Cassit{}~UVIS \citep{Colwell06,Colwell07} and \Cassit{}~VIMS \citep{Hedman07,NH09} explain the observations through the use of simple models that assume an optical-depth dichotomy, with nearly-opaque wakes (with optical depth $\tau_{wake}$) and a low but relatively constant optical depth in the spaces between the wakes ($\tau_{gap}$). However, while some preliminary work has shown that such a model can produce brightness profiles that are consistent with data \citep{Hedman07,SaloLondon08}, it has not been verified that such a bi-modal model describes the actual nature of simulated wakes, not to mention real ones. And if it does, how do $\tau_{wake}$ and $\tau_{gap}$ relate to environmental parameters such as overall occultation optical depth and coefficient of restitution?  What do the values of $\tau_{gap}$ inferred from observations tell us about the conditions under which wakes occur?  

\citet{Salo04} and \citet{Porco08}, on the other hand, produce plots of ring brightness at various geometries for various simulation input parameters, inferring the best input parameters by finding simulated brightnesses that best compare to observations.  But where is the brightness coming from?  For images in reflected light, is the brightness varied primarily by changes in the opacity of dense wake structures, or by the fractional area covered by them?  For images in transmitted light, do more photons get scattered into the camera from the rarefied regions between the wakes (the wakes themselves being largely too opaque to allow much light to pass through them), or are more photons being scattered by the edges of the wake structures?  Better understanding of these questions would help to guide future simulations, and might allow one to infer more from the available observational data.

Yet another current question concerns the nature of ``propeller'' structures observed in images of the mid-A ring \citep{Propellers06,Propellers08,Sremcevic07}, each of which is inferred to be a disturbance surrounding an unseen moonlet embedded within the ring.  Observed propellers generally show the disturbed region as brighter than the background ring brightness, though theoretical models \citep{SS00,SSD02,Seiss05,LS09} indicate that the most prominent feature of the disturbance is a decrease in local density.  Among other hypotheses for this curious behavior, \citet{Propellers06,Propellers08} have proposed that self-gravity wakes tend to lock up ring material into a photometrically inactive state, and that propellers can release this material by locally and temporarily disrupting the wake structure, thus causing the disturbed area to contain more photometrically active material even if it contains less material overall.  Our parameterization of self-gravity wakes allows us to evaluate this hypothesis in a more quantitative fashion than has previously been possible.

A major reason why these questions remain largely unanswered lies in the difficulty of \textit{local density estimation}, the quantification of an underlying density function based on the locations of a finite number of particles, for a strongly heterogeneous density distribution.  The simplest method of density estimation is to divide a simulation space into bins of equal size and to count the number of particles in each bin, but this fails to give meaningful results for current simulations of self-gravity wakes.  This is because the rarefied regions require large bins in order to accumulate enough particles in each bin to give good statistics, but such large bins will badly smear the sharp boundaries between the dense wakes and the rarefied regions.  Contrariwise, if the bins are small enough to resolve the sharp boundaries, nearly all bins in the rarefied regions will contain zero particles (with a few containing one particle), giving no sense of the average density in the rarefied regions. 

We have constructed a density estimation method, to be applied to self-gravity wake simulations, based on circular bins that expand in rarefied regions and contract in dense regions.  A given circular bin is only used if the particles within it satisfy certain criteria (described in detail below) to ensure that they are sufficiently evenly distributed; if those criteria are met, then the bin grows until they are no longer met.  The overlapping circular bins are then projected onto a finely-meshed rectangular grid; nearly all bins in the latter grid contain multiple overlapping circular bins, and the value of the density at each location is taken from the average of the overlapping circular bins.  

Applying our density estimation method to a set of numerical simulations of a ring patch characterized by self-gravity wakes, we parameterize the general properties of each simulation as a weighted combination of three optical depth values.  This semi-analytic treatment allows us to comment on trends that are applicable to a wide range of input conditions.  We specifically discuss the implications of our results for previous interpretations of occultation data, as well as for the question of why ``propellers'' appear as relative-bright features. 

Section~\ref{WakeSims} describes our suite of numerical simulations, the results of which constitute the input data for our semi-analytic method.  Section~\ref{DensEst} describes our density estimation method.  Section~\ref{Results} describes our results, and Section~\ref{Discussion} provides further discussion.  Section~\ref{Conclusions} presents a summary and conclusions. 

\section{Wake Simulations \label{WakeSims}}

\begin{table}[!b]
\begin{scriptsize}
\caption{Parameters for our simulations of self-gravity wakes. \label{WakeSimsTable}}
\begin{tabular} { c c c c c }
\hline
\hline
$\tau_{dyn}$ & $\sigma$ & $N$ & $\lambda_{cr}$ & Coefficient of restitution law \\
\hline
0.1 & 10 g~cm$^{-2}$ & 7,359 & 15.3~m & Borderies, $v^* = 0.001$~cm~s$^{-1}$ \\
0.2 & 20 g~cm$^{-2}$ & 14,718 & 30.5~m & Borderies, $v^* = 0.001$~cm~s$^{-1}$ \\
0.3 & 30 g~cm$^{-2}$ & 22,077 & 45.8~m & Borderies, $v^* = 0.001$~cm~s$^{-1}$ \\
0.35 & 35 g~cm$^{-2}$ & 25,757 & 53.4~m & Borderies, $v^* = 0.001$~cm~s$^{-1}$ \\
0.4 & 40 g~cm$^{-2}$ & 29,436 & 61.0~m & Borderies, $v^* = 0.001$~cm~s$^{-1}$ \\
0.45 & 45 g~cm$^{-2}$ & 33,116 & 68.6~m & Borderies, $v^* = 0.001$~cm~s$^{-1}$ \\
0.5 & 50 g~cm$^{-2}$ & 36,796 & 76.3~m & Borderies, $v^* = 0.001$~cm~s$^{-1}$ \\
0.5 & 50 g~cm$^{-2}$ & 36,796 & 76.3~m & Bridges \\
\hline
\end{tabular}
\\ $\tau_{dyn}$ is input mean dynamical optical depth, $\sigma$ is input mean surface density, 
\vspace{-0.08in}
\\ $N$ is total number of particles, and $\lambda_{cr} = 4 \pi^2 G \sigma / \kappa^2$ is the Toomre critical wavelength
\end{scriptsize}
\end{table}

We carried out a series of numerical simulations of an orbiting patch of ring particles.  Details of the numerical technique are provided in \citet{Porco08}, but are briefly summarized here for convenience.  Particle trajectories are computed in a rotating frame (the patch) using Hill's equations of motion with self-gravity.  The orientation of the patch is such that the $x$-direction points radially away from the planet, the $y$-direction is in the direction of the patch center motion, and $z$ is perpendicular to $x$ and $y$ according to the right-hand rule. Boundary conditions are applied in the $x$ and $y$ directions to keep the particle number constant.  Duplicates of the patch surround it in the orbital plane to provide a smoother gravity potential and to allow for collisions at the patch boundary.  Duplicates in the $\pm x$ directions are offset in $\pm y$ to account for differential shear \citep{WT88}.  A second-order leapfrog integrator adapted to the rotating frame is used to integrate the equations of motion.  Collisions are predicted during the ``drift'' update of each integration step and are carried out using billiard-ball restitution equations.  We ignored surface friction and particle spin.  We chose a timestep of 5~seconds, which is more than 1,000 times smaller than either the orbital period or the dynamical interaction time between two particles, $\sqrt{1/G \rho}$.  Initial conditions consist of a uniform distribution of dynamically cold particles in a thin slab; equilibrium, measured as a flattening of the components of the velocity dispersion in the patch, is established typically within 10 orbits of the central patch around Saturn.

We used seven values of the input mean dynamical optical depth $\tau_{dyn}$, defined as the total cross-section area of particles divided by the total patch area (Table~\ref{WakeSimsTable}).  All simulations used a Saturn-centered orbital distance of 130,000~km, a monodisperse particle-size distribution (i.e.,~identically-sized particles) of radius $R = 1.667$~m, and internal density equal to the local Roche critical density 0.45~g~cm$^{-3}$ \citep{PorcoSci07,Porco08}.  For this combination of particle properties, which are typical values for the mid-A~ring, the numerical values for the input mean surface densities (in units of g~cm$^{-2}$) are simply 100 times the numerical values for $\tau_{dyn}$.  The patch dimensions of $510 \times 1260$~m, with the short axis oriented in the radial direction and the long axis along the orbital direction, were chosen to be always greater than $4 \times 10$ times the Toomre critical wavelength ($\lambda_{cr} = 4 \pi^2 G \sigma / \kappa^2$, where $G$ is Newton's constant and $\kappa$ is the epicyclic frequency), even for the simulations with the highest densities, which ensures that no individual particle or structure can reach across the periodic boundary conditions to meaningfully interact with itself.  

Other investigators have used a parameter $r_p$ \citep{Ohtsuki93,Salo95} or $r_h^*$ \citep{Daisaka01}, a modified ratio of Hill radius and particle radius, to track the susceptibility to wake formation of a given simulated ring patch.  This parameter depends only on the particle's internal density and semimajor axis \citep[\Eqn{}~8]{Daisaka01}.  Furthermore, the Roche critical density, at which a particle fills its own Hill sphere, depends only on the particle's semimajor axis \citep[\Eqn{s}~1 and~2]{PorcoSci07}.  Combining these two, any particle at the Roche critical density will have $r_h^* \sim 0.85$.  One can now locate our suite of simulations as a vertical line in the two-dimensional parameter space shown in \Fig{}~2 of \citet{Daisaka01}, and can calculate from their \Eqn{}~13 that the onset of self-gravity wakes should occur for $\tau \gtrsim 0.25$.

For the primary batch of simulations, we used the coefficient of restitution stated by \citet{BGT84}, building on the work of \citet{Andrews1930}, with $v^* = 0.001$~cm~s$^{-1}$.  This relatively dissipative law, which yields lower post-collision speeds than the more commonly-used coefficient-of-restitution law formulated by \citet{Bridges84}, was identified in the analysis of \citet{Porco08} as leading to a better fit to observations of this particular region of the A ring.  Furthermore, we carried out one additional simulation that was identical to the one described above with $\tau_{dyn} = 0.5$ and $\sigma = 50$~g~cm$^{-2}$, but using the \citet{Bridges84} law.  The latter simulation is intended to be identical to that used by \citet{Salo04} for their photometric analysis.  Throughout this paper, comparison of the results given by the two coefficients of restitution can be thought of as giving a crude approximation of the variation that might arise in our models due to uncertainty of various input parameters.

\section{Density Estimation Method \label{DensEst}}

\begin{figure}[!t]
\begin{center}
\includegraphics[width=2.6cm,keepaspectratio=true]{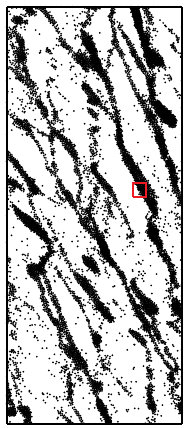}
\includegraphics[width=6cm,keepaspectratio=true]{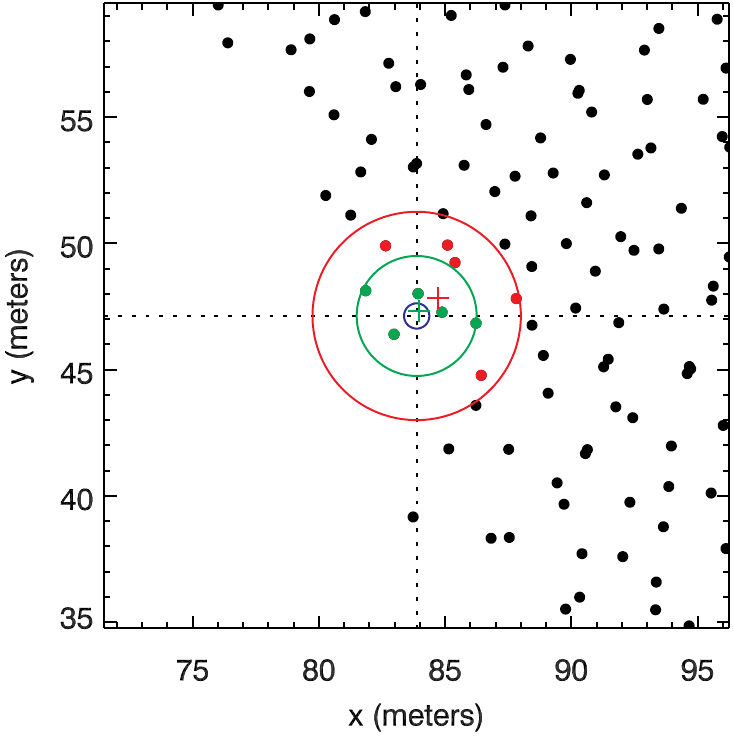}
\includegraphics[width=10.5cm,keepaspectratio=true]{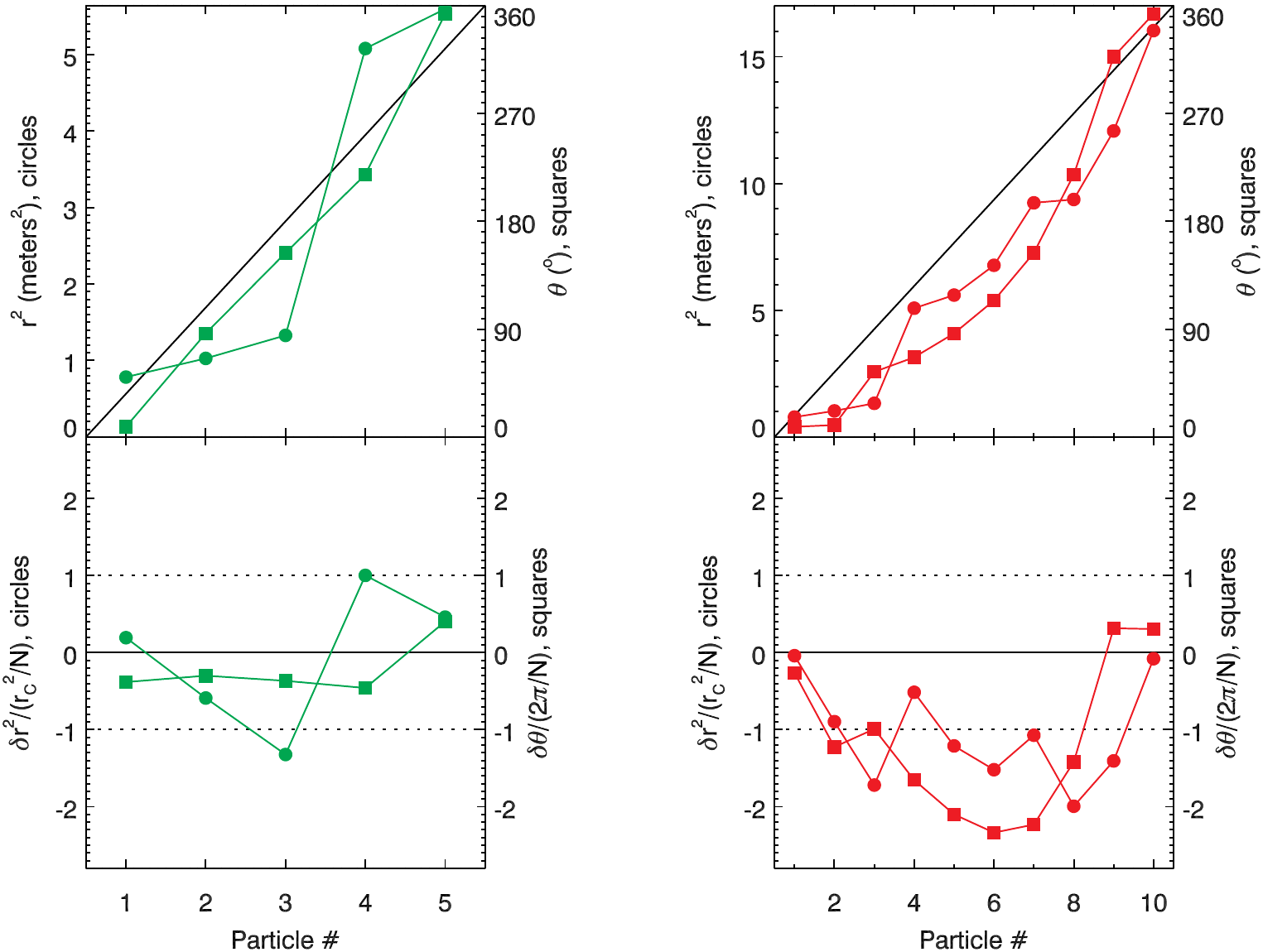}
\caption{Illustration of our density estimation method using a sample snapshot from the simulation with $\sigma=50$~g~cm$^{-2}$ at $t=30$~orbits (top left), at a location centered at $x_0=84$~meters and $y_0=47$~meters from the frame center (inset at top right).  The first circle (green circle, including particles shown in green) was validated by a ``center of mass'' less than 0.5~m from the micro-bin center (i.e.,~the green cross is inside the blue circle) and also by (lower left-hand plots) radial and azimuthal distribution of particles with standard deviations not too far from uniformity.  The attempt to expand the circle (red circle, including particles shown in red and green) failed due to a ``center of mass'' too far from the micro-bin center (i.e.,~the red cross is outside the blue circle), and also due to (lower right-hand plots) excessive deviation from uniformity in both the radial and azimuthal distribution of particles.  In the plots, circles indicate the distribution of $r^2$ values and squares indicate $\theta$ values.  \label{dcr_contour7_example}}
\end{center}
\end{figure}

After each simulation ran for a long-enough period to reach equilibrium (i.e.,~all components of the velocity dispersion have settled to roughly constant values), we took seven snapshots at $t = \{ 18, 20, 22, 24, 26, 28, 30 \}$~orbits.  For each snapshot, containing the positions of all particles at a given time, we performed the following algorithm on a central sub-patch\fn{Our reason for focusing on a sub-patch was simply to reduce computation time.} of dimensions $300 \times 750$~m. 

We first projected the three-dimensional particle positions onto a plane.  We used the ringplane ($z=0$) for this purpose, though this need not be the case.  In order to quantify the distribution of densities observed at a slant-path through the ring, any plane perpendicular to a desired line-of-sight may be used. 

We divided the sub-patch into ``micro-bins'' of size 0.25~m, small enough to clearly resolve the boundaries between wake and inter-wake regions, forming a grid with dimensions $1200 \times 3000$ micro-bins.  At each micro-bin location $[ x_0, y_0 ]$, we begin by drawing a circle around the five particles closest to that location.  The radius of the circle we denote as $r_C$, and each of the $N$ particles within the circle is assigned a position $[ r, \theta ]$ relative to the center of the circle and the $x=0$ axis.  We then perform three tests to determine whether the particles are evenly distributed (\Fig{}~\ref{dcr_contour7_example}):  
\begin{enumerate}
\item{} The average position of the particles inside the circle (the ``center of mass''\fn{This is a figure of speech; in the case of a particle-size distribution, when the average position of particles and the actual center of mass may not be the same, it is particle locations with which we are concerned.}) must be within 0.5~meters of the center of the micro-bin (the blue circle in \Fig{}~\ref{dcr_contour7_example});
\item{} The number of particles whose distance from the center of the circle is less than or equal to $r$ should increase linearly with $r^2$.  That is, for a perfectly even distribution, the expected value of the radial location of the $n$th particle would be $r_n^2 = r_C^2 (n-1/2) / N$, where $n$ ranges from 1 to~$N$.  We compute the standard deviation of the residual between the actual and ideal values of $r_n$, which must be less than the total range divided by the number of particles ($r_C^2/N$) to pass the test.  
\item{} Similarly, for a perfectly even distribution, the angular separation between adjacent particles would be constant, which is to say that the expected value of the longitude of the $n$th particle would be $\theta_n = 2 \pi (n-1/2) / N$.  We find the standard deviation of the residual between the actual and ideal values of $\theta_n$, which must be less than the total range divided by the number of particles ($2\pi/N$) to pass the test.  
\end{enumerate}

If all three tests are passed, then the micro-bin is validated and used in the final solution.  Furthermore, we try to expand the circle, increasing $r_C$ to take in the next five particles and redoing the above-mentioned tests on the new distribution.  If the tests are passed again, the iteration continues until the maximum value of $r_C = 12.5$~meters is reached.  If the tests are not passed, then we revert back to the last successful value of $r_C$ for that location. 

We identify five parameters in our method that require human input.  These include the number of particles added to the circle in each iteration (we chose five, which is large enough for each iteration to have reasonable statistics, but small enough to allow the validation of bins that are close to a wake/inter-wake boundary), the maximum value of $r_C$ (we chose 12.5~meters, which is less than the characteristic spacing between the dense wake structures), and the threshold criteria for the three tests enumerated above (the values described could reasonably be multiplied by a scalar of order unity).  Aside from this human input, our method is fully automated. 

This process is illustrated in \Fig{}~\ref{dcr_contour7_example}, where the first circle (green) results in a successful validation, while the attempt to expand the circle (red) fails.  In the top-right panel, Test \#1 can be seen in that the average position of the first five particles (the green cross) is inside the blue circle, while the average position of the first ten particles (the red cross) is not.  Tests \#2 and \#3 are illustrated in the plots, where the horizontal dotted lines indicate the standard deviation criteria.  Note that the plotted residuals for $r^2$ and $\theta$ have been normalized by the threshold criterion values; when the circle is expanded, the threshold criterion for $\theta$ decreases simply because $N$ is twice as large, while the threshold criterion for $r^2$ slightly increases because a larger $r_C^2$ compensates for the increased $N$. 

In the gaps between the dense wake structures, it commonly happens that even the maximum-size circle ($r_C = 12.5$~meters) contains fewer than five particles.  In this case, we designate the location as ``sparse'' and proceed with the particles that do occur in the maximum-size circle.  For ``sparse'' bins, we waive Test \#1 (the ``center-of-mass'' test), in order to avoid large areas of the inter-wake gaps having no coverage at all with valid bins. 

When the above process is complete, each location on our $1200 \times 3000$ grid has one of two states:  a valid circle of radius $r_C$ with a certain number of particles within it (which can be resolved into a density of number of particles per unit area), or no valid result.  We then visualize the valid circles and find that nearly every location within the grid is covered by one or more overlapping circles.  Finally, we calculate the density at each location on the grid by summing the particles in all circles overlapping that location and dividing by the sum of the areas of the same overlapping circles.\fn{Rather than simply averaging the densities in the overlapping circles, this method weights the result somewhat towards those circles that have more particles, and thus better statistics.}  

\section{Results \label{Results}}

\begin{figure}[!t]
\begin{center}
\includegraphics[width=16cm,keepaspectratio=true]{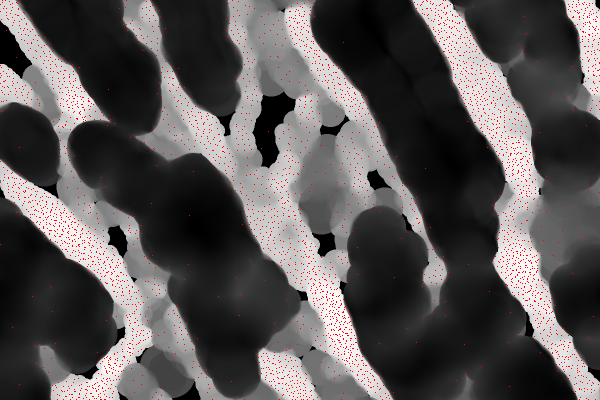}
\caption{Sample result of our density estimation method, showing a portion of the same snapshot shown in \Fig{}~\ref{dcr_contour7_example}.  Red dots are particle locations, which constitute the inputs for density estimation.  Calculated densities are shown as a grayscale background ranging from black (low density) to white (high density).  \label{dcr_contour7_example4}}
\end{center}
\end{figure}

A sample result of our density estimation method is shown in \Fig{}~\ref{dcr_contour7_example4}.  We find that regions with high calculated density (bright in the figure) correlate well with regions with tightly-clustered particles, and that regions with low calculated density (dark in the figure) correlate well with regions in which particles are sparse.  Although some instances of intermediate calculated density (gray in the figure) are on the boundaries between dense and sparse regions, a smearing effect that our method was designed to avoid, we find that the large majority of regions with intermediate calculated density genuinely have particles clustered to an intermediate degree. 

\begin{figure}[!t]
\begin{center}
\includegraphics[width=16cm,keepaspectratio=true]{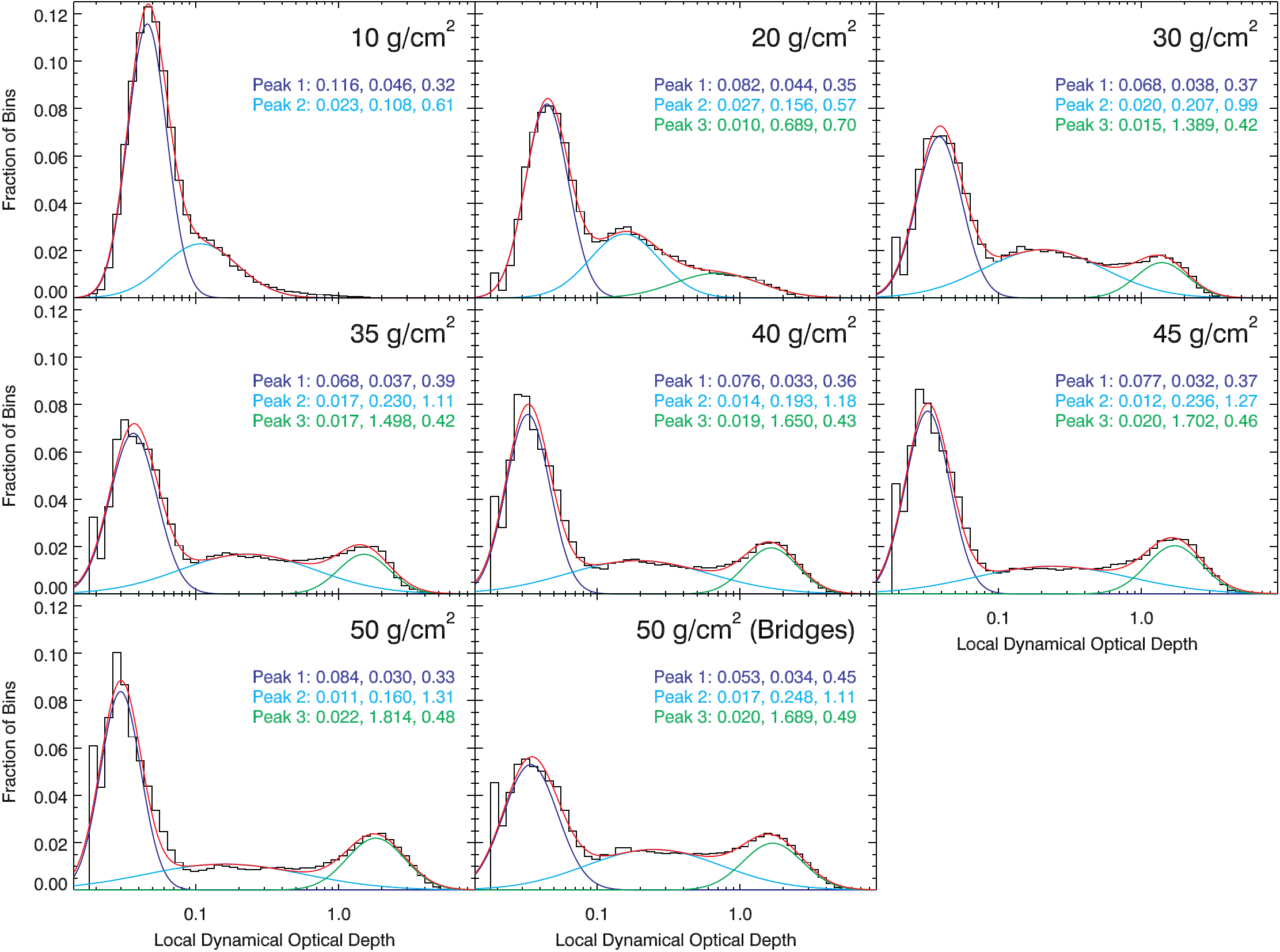}
\caption{Histograms of local dynamical optical depth, $\tau_{dyn}(x,y)$, for the eight simulations described in Section~\ref{WakeSims}, calculated using the method described in Section~\ref{DensEst}.  Least-squares fits are made to a sum of three gaussians (or, in the top-left case, two gaussians).  Within each plot are listed the fit parameters: the height, center location, and full-width at half-maximum (the latter in units of $\log \tau$) of each gaussian.  \label{dcr_histfit4}}
\end{center}
\end{figure}

Now that we have a grid of local densities within a self-gravitating ring patch, we can make a histogram showing the frequency with which each value occurs (\Fig{}~\ref{dcr_histfit4}).  For this purpose, we express the local density in terms of the local dynamical optical depth, $\tau_{dyn}(x,y) = \rho(x,y) \pi R^2 / \ell^2$, where $\rho(x,y)$ is the number density (particles per bin) calculated by the method described in Section~\ref{DensEst}, $\ell = 0.25$~m is the bin size, and $R = 1.667$~m is the particle radius.  Note that this is not the same quantity as that measured in optical observations, the photometric optical depth $\tau_{phot} = -\ln T$, where the transparency $T$ is the fractional area not blocked by particles.  The two are nearly equal at low values, but diverge at large optical depths when particles come together closely enough that their inability to occupy the same location in space becomes important. 

Some of the histograms in \Fig{}~\ref{dcr_histfit4} have an anomalously high value in the left-most bin, containing the lowest optical depths.  This is due to a small fraction of bins which, in the application of our density estimation method, were not overlapped by any valid circular bins.  This generally occurs in small regions of very low density that are surrounded by regions of higher density; such regions are visible in \Fig{}~\ref{dcr_contour7_example4} as uniformly black.  Rather than further refine the method to better account for these difficult cases, we simply assigned to these locations a density equal to the lowest densities obtained.  The fact that these left-most spikes are not very large in any of the histograms justifies our decision to neglect this effect as a small perturbation.

We obtain quite good results by modeling the histograms in \Fig{}~\ref{dcr_histfit4} (using $\log \tau_{dyn}$ as the independent variable) as a sum of two or three gaussians.  We justify the choice of a gaussian fit as follows:  The results of our density estimation, though quite good, do contain some variation within regions that should have a single density; sometimes the local density is higher in the immediate vicinity of a single particle, and sometimes the density decreases smoothly with distance from a group of particles.  If we provisionally assume that the overall density distribution is characterized by a small number of discrete values that each represent a certain fraction of the total area, an assumption that comports with the qualitative sense of many investigators and that forms the foundation of the bimodal-distribution assumption of \citet{Colwell06,Colwell07} and \citet{Hedman07}, then it is reasonable for the variation we just described to transform the density histogram (\Fig{}~\ref{dcr_histfit4}) from a distribution containing only a small number of non-zero values to a distribution characterized by a small number of gaussians.  

\begin{figure}[!t]
\begin{center}
\includegraphics[width=7cm,keepaspectratio=true]{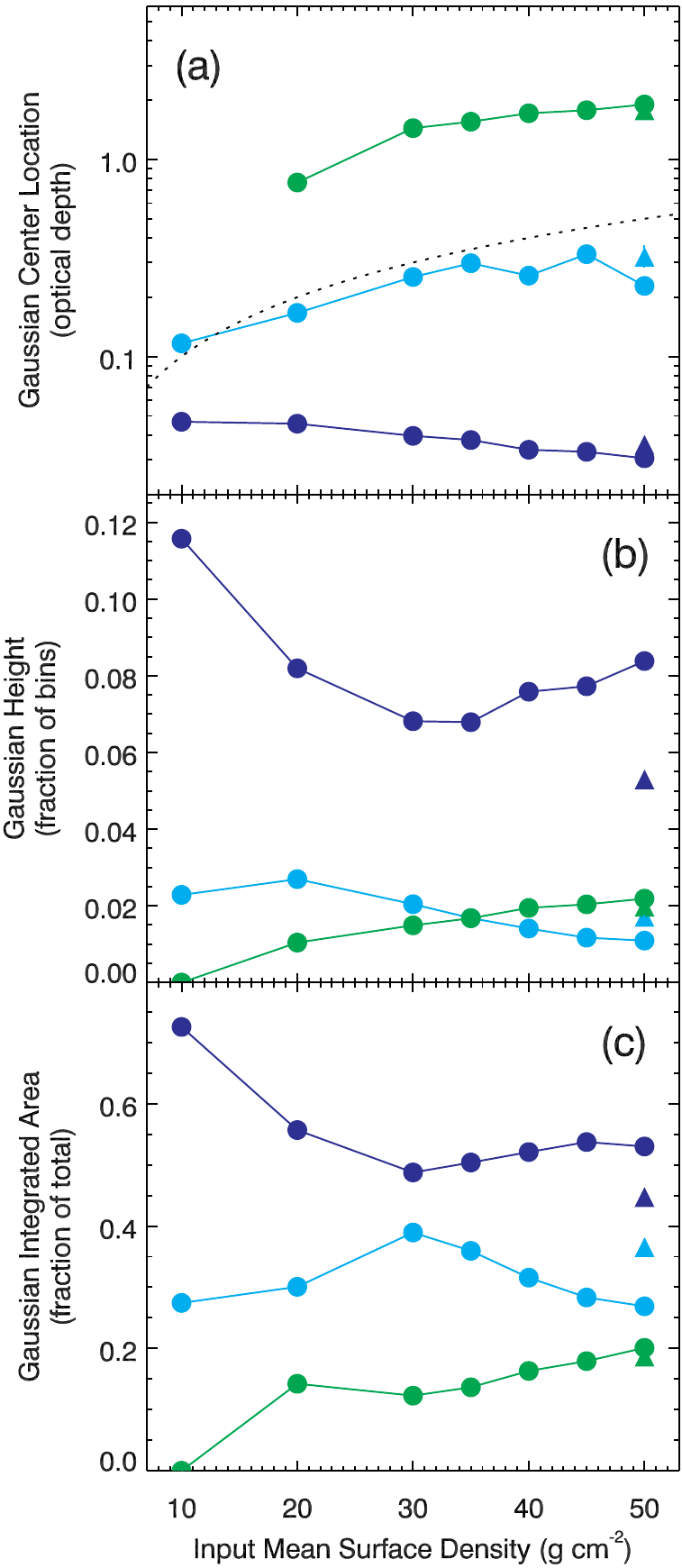}
\caption{Parameters of fitted gaussians from \Fig{}~\ref{dcr_histfit4}.  Expressions are given in the text for (a) the center location, (b) the height, and (c) the integrated area normalized by the total area under the combined fit curve.  The dotted line in panel (a) shows the input mean dynamical optical depth, calculated by dividing the sum of the cross-section areas of all particles by the total area.  The colors of each curve correspond to the colors of the fitted gaussians in \Fig{}~\ref{dcr_histfit4}.  Circles signify the simulations using a \citet{BGT84} coefficient of restitution law with $v^* = 0.001$~cm~s$^{-1}$, and triangles signify the simulation using a \citet{Bridges84} law.  \label{dcr_histfit4a}}
\end{center}
\end{figure}

Defining a single gaussian curve as $y(x) = a_0 e^{-(x-a_1)^2/2a_2^2}$, where $x$ here is $\log \tau_{dyn}$, the parameters of each gaussian as stated in \Fig{}~\ref{dcr_histfit4} are the height of the gaussian in units of the histogram ($=a_0$), the location of the center of the gaussian in units of $\tau_{dyn}$ ($=10^{a_1}$), and the full-width at half-maximum of $\log \tau_{dyn}$ ($=2 \sqrt{2 \ln 2} \cdot a_2$).  The fit parameters are plotted in \Fig{}~\ref{dcr_histfit4a} as follows:  the center location ($=10^{a_1 + a_2^2/2}$) is now the characteristic value of $\tau_{dyn}$ for a log-normal distribution, the gaussian height ($=a_0$) is as in \Fig{}~\ref{dcr_histfit4}, and the integrated area under the gaussian curve ($=\sqrt{2\pi} \cdot a_0 a_2$) is normalized by the total area under the combined fitted curve.

The close correspondence between the density histograms and our sum-of-three-gaussians models justifies \textit{a~posteriori} our provisional assumption.  Surprisingly, though, the density distribution is not bimodal but (except for the case with the lowest surface density) trimodal!  Once the wakes are fully developed (i.e., for surface densities $\sigma \gtrsim 30$~g~cm$^{-2}$ in our simulations), the peak at high optical depth remains relatively stationary with a central value $\tau \sim 2$, and both the central value and the integrated area of the gaussian gradually increase with $\sigma$ (green in \Fig{}~\ref{dcr_histfit4a}).  The largest fraction of the area is covered by very low optical depths, with fitted central values $\tau \sim 0.04$ that gradually decrease with increasing $\sigma$ (dark~blue in \Fig{}~\ref{dcr_histfit4a}); these optical depths are so low that such regions are likely to be photometrically inactive in all cases.  Optical depths closer to the $\tau_{gap}$ values inferred by previous authors characterize the middle peak in each histogram, with central values ranging from $\tau \sim 0.1$ to $\tau \sim 0.25$ (cyan in \Fig{}~\ref{dcr_histfit4a}).  

This result leads us to suggest re-interpreting the results of previous investigators \citep{Colwell06,Colwell07,Hedman07,NH09}, whose models assume that a photometrically active $\tau_{gap}$ characterizes the entire area that is not occupied by the dense wakes.  As we will show in Section~\ref{Occultations}, our results are likely consistent with this picture insofar as $\tau_{gap}$ is interpreted simply as the area-weighted average optical depth over the inter-wake regions; however, our results indicate that most of that area may be so sparsely populated that it is not photometrically active at all, while a smaller fraction of the inter-wake regions contains somewhat higher optical depths that cause the entire inter-wake region to average out to $\tau_{gap}$.  When one looks at movies of our self-gravity wake simulations, these regions of intermediate $\tau$ most commonly arise from formerly dense wakes that are in the process of being disrupted.  

It is possible that the presence in our numerical simulations of large areas with very low optical depth might be a result of our use of a monodisperse size distribution (that is, we used a single particle size, namely $R=1.667$~m).  Preliminary work by \citet{SS07} indicates that smaller particles may be more likely to escape the dense wake and spread into the inter-wake regions.  Although we do not in this paper investigate simulations with a particle size distribution, this could be done easily enough by separating simulated particles into logarithmic size bins, perhaps with widths of a decade or a half-decade, assigning an average size to the particles within each bin, proceeding with a separate density estimation for each size bin, and finally combining the resulting estimated densities.  On the other hand, there is considerable evidence that the clearing of small particles out of the inter-wake regions is in fact required in order to match brightnesses derived from $N$-body wake simulations with the amplitude of the azimuthal brightness asymmetry observed in the A~ring in both ground-based and spacecraft imaging data, as first suggested by \citet{PorcoDPS03} and seen in the simulations of \citet{Porco08}.  The clearing out of inter-wake regions occurs in models using a lossier (``squishier'') coefficient of restitution, which \citet{Porco08} argue are more realistic and provide a better fit to observations.  Indeed, in our \Fig{s}~\ref{dcr_histfit4} and~\ref{dcr_histfit4a}, it is clear that the simulation using the ``bouncier'' \citet{Bridges84} coefficient of restitution law yields a much smaller peak (i.e., less area covered) at very low optical depths. 

\begin{figure}[!t]
\begin{center}
\includegraphics[width=8cm,keepaspectratio=true]{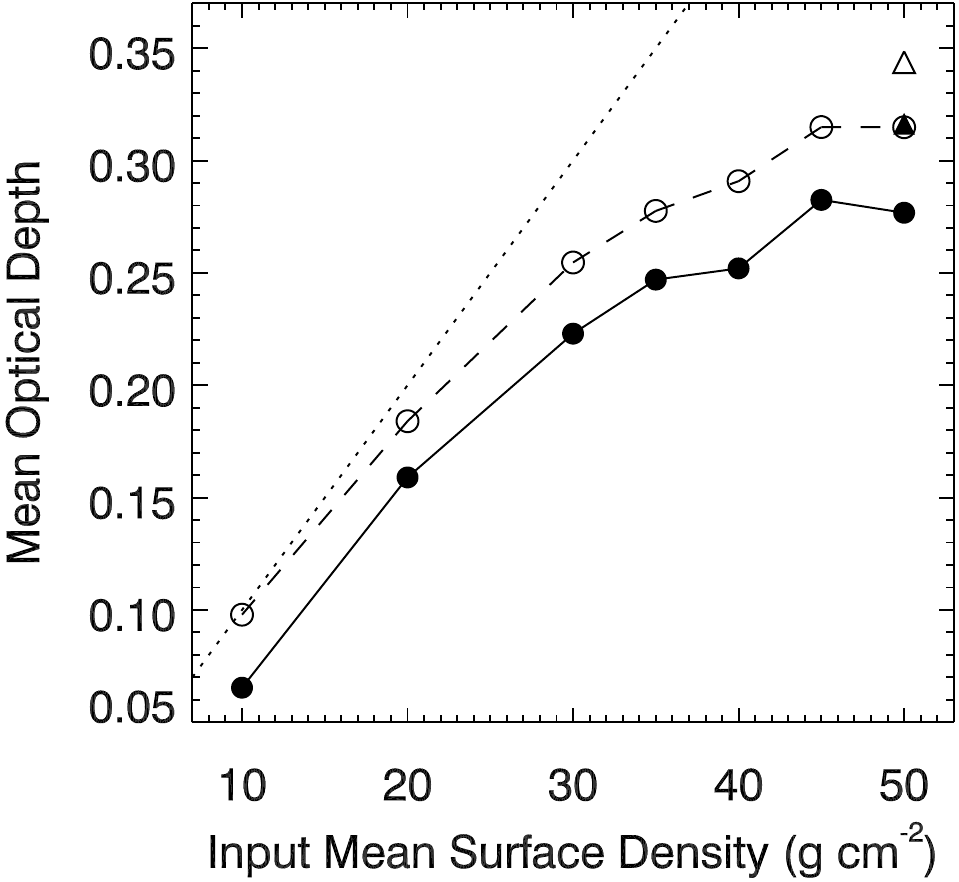}
\caption{Solid symbols connected by solid line are parameterized mean optical depths, calculated by combining the central optical depths of each gaussian peak (\Fig{}~\ref{dcr_histfit4a}a) weighted by their relative integrated areas (\Fig{}~\ref{dcr_histfit4a}c).  Open symbols connected by dashed line are actual mean photometric optical depth, calculated by dividing the area blocked by particles by the total area.  Dotted line shows the input mean dynamical optical depth, calculated by dividing the sum of the cross-section areas of all particles by the total area.  In all cases, circles signify the simulations using a \citet{BGT84} coefficient of restitution law with $v^* = 0.001$~cm~s$^{-1}$, and triangles signify the simulation using a \citet{Bridges84} law.  \label{dcr_histfit4b}}
\end{center}
\end{figure}

As a check on the accuracy of our parameterization methods, in \Fig{}~\ref{dcr_histfit4b} we compare several methods of calculating the total optical depth of a simulated ring patch.  The dotted line is the input mean dynamical optical depth, which would be equal to the actual mean photometric optical depth if particles were randomly distributed.  The open symbols connected by a dashed line indicate the actual mean photometric optical depth, which can be easily calculated for the entire patch (though not locally); the transparency $T$ is simply the fractional area not blocked by particles, measured by projecting particle positions onto a plane perpendicular to the line of sight, and $\tau_{phot} = - \ln T$.  We note in passing that our calculated actual mean optical depth for the \citeauthor{Bridges84} simulation is 0.34; in \Fig{}~16 of \citet{Salo04}, whose simulation we intentionally attempted to replicate for purposes of comparison, those authors found a photometric optical depth of 0.36 for $B=90^\circ$. 

Ideally, the calculated total optical depths using our density estimation would be equal to the actual mean photometric optical depth.  The solid symbols connected by a solid line indicate the total optical depth obtained by combining\fn{In general, optical depths are properly combined by converting them to transparencies ($T = e^{-\tau}$), finding the mean of the transparencies weighted by their occurrence frequencies, and then converting the total transparency back into optical depth, thus:
\begin{equation}
\label{CombineOpticalDepth}
\tau_{combined} = -\ln \left[ \frac{1}{N} \sum_{i=1}^N f_i e^{-\tau_i} \right] . 
\end{equation}} the central locations of the gaussian peaks (\Fig{}~\ref{dcr_histfit4a}a) weighted by the integrated area under each gaussian curve (\Fig{}~\ref{dcr_histfit4a}c).  If this final method of expressing the optical depth can be legitimized, then we will have succeeded in truly parameterizing our simulated self-gravity wakes, expressing their photometric properties by means of only a few numbers, whose dependence on surface density and other input factors can then be easily tracked.

Firstly, we verify that the mean optical depth calculated from our parameterization is consistent with the mean optical depth calculated by simply adding together the continuum of density values weighted by the histograms shown in \Fig{}~\ref{dcr_histfit4}.  These two are consistent within $\Delta \tau \sim 0.003$ (except for the first point, at $\sigma = 10$~g~cm$^{-2}$, for which $\Delta \tau \sim 0.01$), a small variation given that the curve covers a range of $\Delta \tau \sim 0.2$.

Secondly, however, we find that the parameterized mean optical depth (solid symbols) is slightly but uniformly too low compared to the actual mean optical depth (open symbols).  The reason for this, we believe, is that we have heretofore used the dynamical optical depth $\tau_{dyn}$ (which is easily calculated from dynamical simulations), rather than the photometric optical depth $\tau_{phot}$ (which corresponds to observations).  As we have mentioned, these two are roughly equal when particles are randomly distributed; this is because a gaussian probability of particle overlap (as seen when projected onto a plane perpendicular to any given line of sight) plays the same role in $\tau_{dyn}$ that the exponential plays in $\tau_{phot}$.  However, $\tau_{phot} > \tau_{dyn}$ when the distance between particles is comparable to the particle size (i.e., at a high volume filling factor) because particles are now constrained as to the locations in space they can occupy.  

\begin{figure}[!t]
\begin{center}
\includegraphics[width=7cm,keepaspectratio=true]{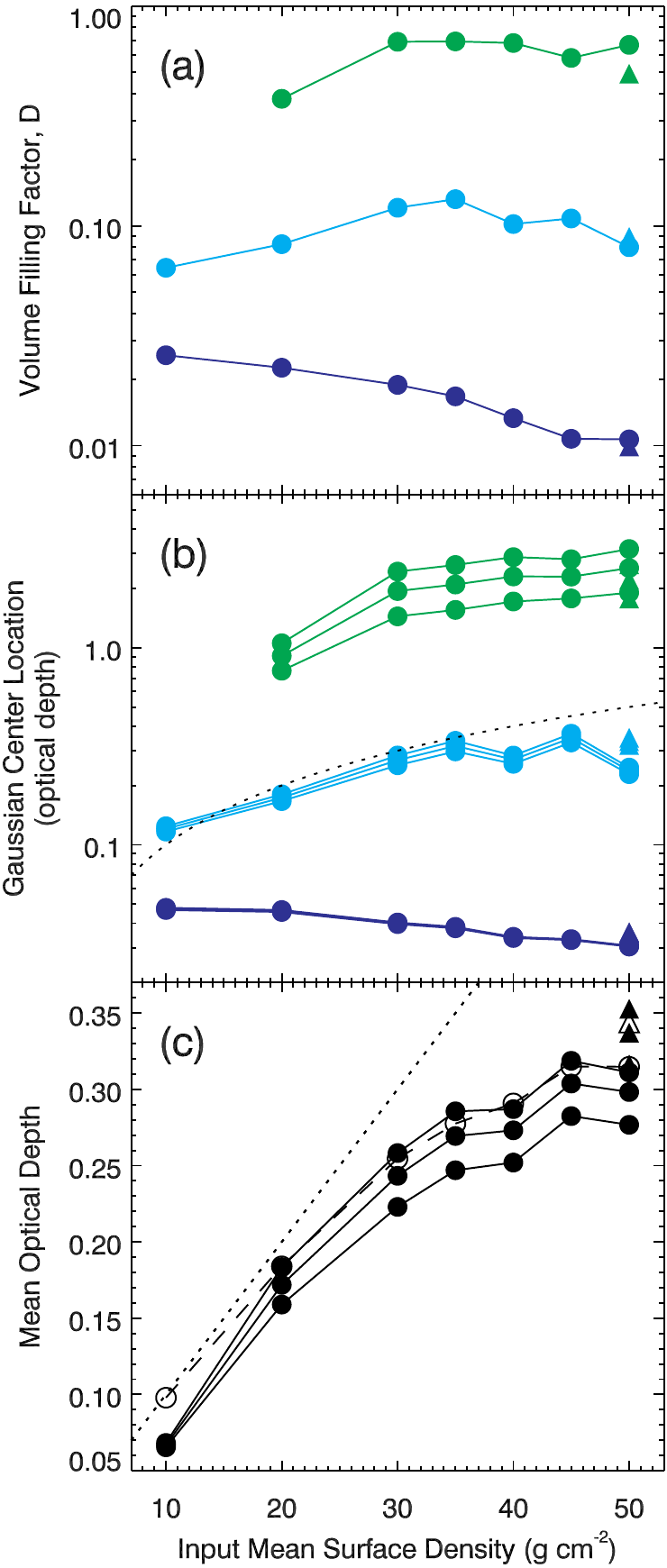}
\caption{(a) Volume filling factor, $D$, calculated from \Eqn{}~\ref{fillingfactor} for each parameterized peak using values of $R$ and $H$ obtained for each simulation as a whole, and using the peak center location as $\tau_{dyn}$.  (b) Revised peak locations (compare \Fig{}~\ref{dcr_histfit4a}a) after applying \Eqn{}~\ref{tauconv}, using $k$ values of 0, 0.5, and~1.  (c) Revised mean parameterized optical depth (cf.~\Fig{}~\ref{dcr_histfit4b}), calculated by combining the revised central optical depths of each gaussian peak (previous panel) weighted by their relative integrated areas (\Fig{}~\ref{dcr_histfit4a}c).  \label{dcr_histfit4c}}
\end{center}
\end{figure}

\citet{SK03} analyzed the photometric properties of simulated particle disks, finding that 
\begin{equation}
\label{tauconv}
\tau_{phot}/\tau_{dyn} \simeq 1 + kD , 
\end{equation}
\noindent where $k$ is a scalar of order unity and 
\begin{equation}
\label{fillingfactor}
D = (4R/3H) \tau_{dyn}
\end{equation}
\noindent is the volume filling factor for vertical scale-height $H$ and particle radius $R$.  It is simple to obtain $H$ from the results of our numerical simulations; it is twice the standard deviation of the $z$-coordinates of the particles, and we find it to range from 4~to 8~m.  We attempted to calculate separate values of $H$ for low-density and high-density bins but did not get a result that was statistically more meaningful than simply using a single value for each simulation; thus we have done the latter.  All of our simulations used $R = 1.667$~m.  

The filling factors for each of our parameterized peaks are shown in \Fig{}~\ref{dcr_histfit4c}a.  Here we have used in place of $\tau_{dyn}$ the center location in optical depth of each peak (\Fig{}~\ref{dcr_histfit4a}a).  We then use the filling factor values to correct the peak locations via \Eqn{}~\ref{tauconv}, as shown in \Fig{}~\ref{dcr_histfit4c}b.  We show results for three values of $k$, namely 0, 0.5, and~1.  The revised mean parameterized optical depth is shown in \Fig{}~\ref{dcr_histfit4c}c.  Now we find that the mean optical depth given by our parameterization corresponds well with that calculated directly, especially for $k=1$, similar to the value of $k$ found to give the best results by \citet{SK03}.  

The uncertainties in the parameter fits from which we obtained optical depth values (\Fig{s}~\ref{dcr_histfit4} and~\ref{dcr_histfit4a}) are difficult to quantify due to the high degree of correlation among the parameters.  To estimate their order of magnitude, we fit two independent gaussians to the two most peak-like regions (namely, $\tau_{dyn} < 0.07$ and $\tau_{dyn} > 1$) of a characteristic histogram (namely, that with $\sigma = 45$~g~cm$^{-2}$).  We found, under the assumption that we had chosen the proper fitting function, that the parameters $a_0$, $a_1$, and $a_2$ were robust within a few percent.

\section{Discussion \label{Discussion}}
We have now successfully parameterized our simulated self-gravity wakes, expressing their properties in terms of three optical depths that are weighted by the integrated area under each gaussian curve.  After converting from dynamical to photometric optical depth by estimating the volume filling factor and using $k=1$ for the scalar in \Eqn{}~\ref{tauconv}, we find that the mean optical depth calculated through our parameterization corresponds well to that calculated directly from the simulations.  We now proceed to apply this method to analysis of \Cassit{} data from occultations and from imaging. 

\subsection{Occultations \label{Occultations}}
In contrast to the bimodal optical-depth models employed by previous analysis of occultation data \citep{Colwell06,Colwell07,Hedman07,NH09}, we find for a ring patch in the mid-A~ring with well-developed self-gravity wakes that slighly more than half of the area is taken up by space that for practical purposes is completely empty and photometrically inactive.  The dense wakes themselves take up slightly less than one-fifth of the area, and the remaining one-fourth to two-fifths is characterized by the so-called $\tau_{gap}$.  Although simulations more detailed and accurate than those we performed might result in some quantitative changes to these results, the general idea of a trimodal distribution of optical depth in the rings deserves consideration.  As we will show, it turns out that the two models perform nearly identically except at very low values of the ring opening angle $B$, which ranges from zero (seeing the rings edge-on) to 90$^\circ$ (seeing the rings face-on).

A trimodal optical depth distribution should primarily affect the estimations of the gap optical depth $\tau_{gap}$ and the gap filling fractions $F_i$ for the simplified wake models.  To explore this issue, let us consider the transmission through the rings as a function of ring opening angle $B$ in the case where the projection of the line of sight to the star onto the ringplane is parallel to the mean wake orientation, which was the geometry studied by \citet{NH09}.  In the formulation used by those authors, the fraction of the visible area taken up by inter-wake regions $F_2$ is equal to their parameter $G/\lambda$, while $F_1 = 1-F_2$ is ignored because $\tau_{wake}$ is taken to be infinite for practical purposes.  Thus, for the simplified model of opaque wakes separated by gaps of constant optical depth (that is, the bimodal case) the transmission is given by
\begin{equation}
T^b(B) = F_2 e^{ - \tau_{gap} / \sin B}.
\end{equation}
Similarly, for a model with a trimodal distribution of optical depths, the transmission is
\begin{equation}
\label{TransmissionTrimodal}
T^t(B) = \sum_{i=1}^3 F_i e^{-\tau_i/\sin B} .
\end{equation}
where $F_i$ and $\tau_i$ are the area fractions and optical depths of the three components of the ring (e.g.,~\Fig{s}~\ref{dcr_histfit4a}c and~\ref{dcr_histfit4c}b, respectively). 

As discussed by \citet{NH09}, it is useful to consider the apparent normal optical depth, 
\begin{equation}
\label{NormalTauGeneral}
\tau_n = - \sin B \ln T , 
\end{equation}
\noindent For the simplified (bimodal) wake model, this is a linear function of $\sin B$: 
\begin{equation}
\label{NormalTauBimodal}
\tau_n^b = \tau_{gap} - \ln F_2 \sin B.
\end{equation}
By contrast, the more complex trimodal model cannot be expressed so simply and will therefore yield a $\tau_n^t$ that is a nonlinear function of $\sin B$.

\begin{figure}[!t]
\begin{center}
\includegraphics[width=7cm,keepaspectratio=true]{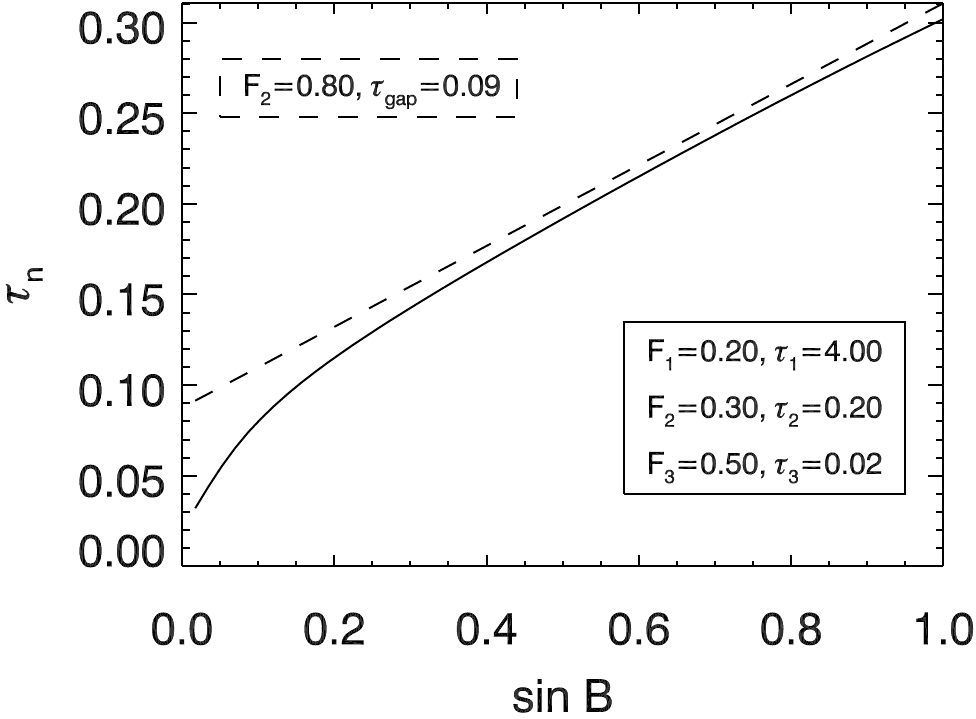}
\caption{Apparent normal optical depth $\tau_n$ as a function of the sine of the ring opening angle $B$ for a bimodal (dashed) and a trimodal (solid) model.  The bimodal model is calculated from \Eqn{}~\ref{NormalTauBimodal} with $F_2 = 0.8$ and $\tau_{gap} = 0.09$, while the trimodal model is calculated from \Eqn{s}~\ref{TransmissionTrimodal} and~\ref{NormalTauGeneral} with $F_i = \{ 0.2, 0.3, 0.5 \}$ and $\tau_{gap} = \{ 4.0, 0.2, 0.02 \}$, respectively.  Specific values used for the area fractions $F_i$ and optical depths $\tau_i$ are shown in the plot.  \label{tiscmodcomp}}
\end{center}
\end{figure}

Apparent normal optical depths $\tau_n$ for both bimodal and trimodal models are plotted in \Fig{}~\ref{tiscmodcomp}, for values of the area fractions and optical depths roughly consistent with those found in Section~\ref{Results}.  The bimodal model uses a gap fraction equal to the sum of the fractions of the two lower-optical-depth components in the trimodal model and a gap optical depth equal to the area-weighted mean optical depth in the two lower-optical-depth regions.  For intermediate to moderately high ring opening angles, these two curves match reasonably well (up to a slight offset), which indicates that the optical depth of gaps derived from the occultation data is the mean optical depth of the regions outside the optically thick wakes. Some slight deviation from a straight line can be observed at the largest opening angles ($\sin B \simeq 1$), as the occultations become marginally sensitive to the finite optical depth of the wakes.  A more dramatic deviation from the linear trend, however, can also be seen at very low opening angles ($\sin B \lesssim 0.2$, which is to say $B \lesssim 10^\circ$), where the moderate-optical-depth regions start to appear opaque, and thus the occultations begin to become capable of distinguishing between the very-low- and moderate-optical-depth components of the model. 

Thus, occultations at very low opening angles may be able to discern variations in the optical depths within the gaps.  Otherwise, existing analysis of occultation data should simply be interpreted with $\tau_{gap}$ representing the area-weighted average optical depth within the gaps (or inter-wake regions), keeping in mind the possibility that there may be strong variations in optical depth within those inter-wake regions. 

\subsection{Imaging \label{Imaging}}
Porco \etal{} (2008) have already applied detailed direct photometric modeling, using a deterministic geometric ray-tracing method to shoot photons at simulated ring patches characterized by self-gravity wakes, to match the brightness of the ring as seen in \Cassit{} images.  Our semi-analytic method is not able to improve on their analysis in terms of quantitative properties, but we can speak to some trends that underlie the rings' photometric behavior.

For images of the lit face of the rings, the large majority of the brightness comes from the dense wakes, which reflect more light than do regions of lesser optical depth.  For images of the unlit face of the rings, on the other hand, most of the brightness comes from regions of intermediate optical depth, as the dense wakes are largely opaque and the nearly-empty regions do not scatter much light towards the camera.  It would be instructive for future papers on direct photometric modeling to show at least a sample image of the brightness resulting from a ring patch; we predict that the brightness on the unlit side of the rings will be dominated by a relatively small area of intermediate optical depth, rather than being broadly spread over the inter-wake region.  Furthermore, the trends discussed in this paper can help future direct photometric modeling to better tune their simulation parameters to match measured brightnesses. 

A more direct application of our parameterization is useful in investigating the peculiar photometric properties of so-called ``propellers''.  These local disturbances in the rings are thought to be due to the perturbing influence of embedded $\sim 100$-meter moonlets.  Observed propellers \citep{Propellers06,Propellers08,Sremcevic07} generally show the disturbed region as brighter than the background ring brightness, though theoretical models \citep{SS00,SSD02,Seiss05,LS09} indicate that the most prominent feature of the disturbance is a decrease in local density.  For images of the unlit side of the rings, depending on the background optical depth, brightness sometimes decreases with increasing density as the rings become more opaque, but calculations of the photometric properties based on \citet{Chandra60} indicate that this is not the operating regime for the images in question, given their particular viewing geometry and the mean surface density as measured from spiral density waves \citep{soirings}.  Furthermore, for images of the lit side of the rings, denser regions are always brighter than less dense regions. 

\begin{figure}[!t]
\begin{center}
\includegraphics[width=7cm,keepaspectratio=true]{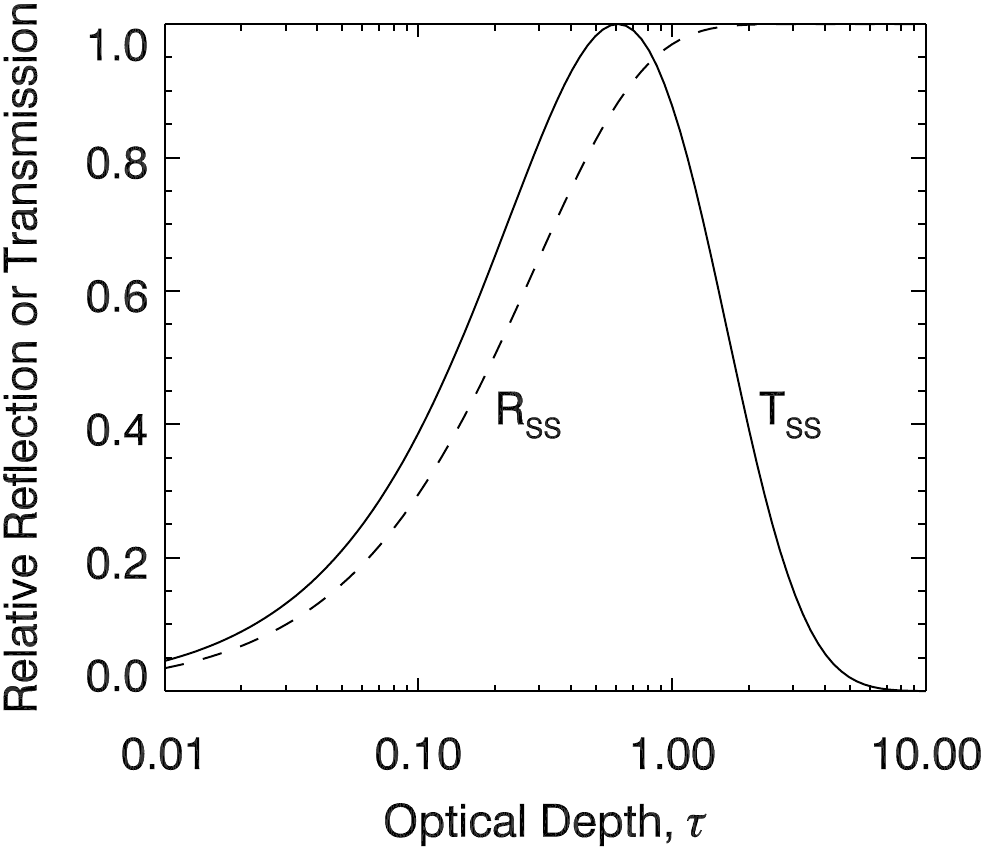}
\caption{Relative single-scattering brightness of a homogeneous slab of particles, calculated from \Eqn{s}~\ref{ChandraR} (reflection, here shown as dashed line) and~\ref{ChandraT} (transmission, here shown as solid line) using $\mu=1$ and $\mu_0=0.4$.  \label{dcr_chandra60}}
\end{center}
\end{figure}

We apply our parameterization of self-gravity wakes to this problem by converting the three optical depth values for each simulation to brightnesses and then performing an area-weighted average as before.  \citet{Chandra60} derived for a homogeneous slab the single-scattering reflection ($R_{SS}$) and transmission ($T_{SS}$): 
\begin{equation}
\label{ChandraR}
R_{SS} \propto 1 - e^{ -\tau/\mu - \tau/\mu_0 }
\end{equation}
\noindent and
\begin{equation}
\label{ChandraT}
T_{SS} \propto e^{-\tau/\mu} - e^{-\tau/\mu_0} ,
\end{equation}
\noindent where $\mu$ and $\mu_0$ are the respective cosines of the emission angle (the angle of the direction towards the camera from the ring-plane normal, which is the complement of the ring opening angle $B$) and the solar incidence angle (the angle of the direction towards the Sun from the ring-plane normal).  We neglect the albedo and the phase function, which do not vary among the instances we consider and thus do not contribute to any trends.  We will use $\mu=1$, corresponding to an image in which the camera is looking straight down onto the face of the rings (which is to say $B = 90^\circ$), because in carrying out our density-estimation method we projected ring particle positions onto the ring plane.  To investigate behavior at other values of $\mu$, one could redo the density estimation with ring particle positions projected onto a plane perpendicular to a different line of sight.  Nevertheless, the general trends illuminated in our analysis here should be applicable to a broader range of viewing geometries.  We use $\mu_0 \sim 0.4$, appropriate to the Sun's illumination of Saturn's rings in 2004 and 2005, when the first images containing propellers were taken.  The resulting values for $R_{SS}$ and $T_{SS}$ are plotted in \Fig{}~\ref{dcr_chandra60}.  

\begin{figure}[!t]
\begin{center}
\includegraphics[width=7cm,keepaspectratio=true]{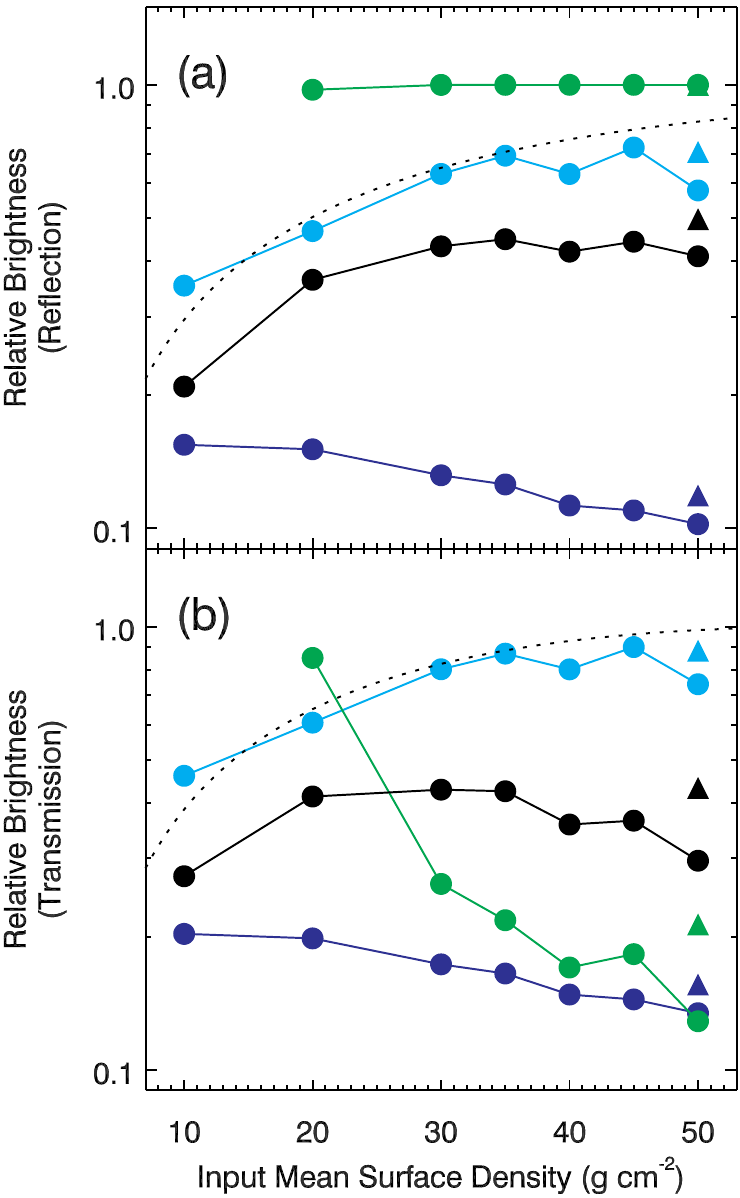}
\caption{Calculated relative brightness for images in (a) reflection and (b) transmission, calculated from the optical depths given in \Fig{}~\ref{dcr_histfit4c}b for $k=1$ using the single-scattering brightnesses shown in \Fig{}~\ref{dcr_chandra60} for a viewing geometry defined by $\mu=1$ and $\mu_0=0.4$.  Colors are for the three peaks, as in \Fig{s}~\ref{dcr_histfit4} and~\ref{dcr_histfit4a}, while black is the combined brightness calculated by averaging the values for the three peaks weighted by the integrated areas given in \Fig{}~\ref{dcr_histfit4a}c.  Dotted lines show the brightness associated with the input mean dynamical optical depth, which corresponds to the brightness a ring patch would have if self-gravity wakes were not present.  \label{dcr_brightness}}
\end{center}
\end{figure}

Our use of the single-scattering approximation is justified for images in reflected light \citep{Dones93} and for self-gravity wakes with their high filling factor \citep{SK03}.  However, multiple scattering can significantly brighten images in transmitted light of non-wake regions if the ring reverts to a many-particle-thick structure \citep{SK03}.  Thus, the dotted line in \Fig{}~\ref{dcr_brightness}b may be even higher than we have shown, which only serves to strengthen the arguments made in the next section.

\subsection{Why propellers are bright \label{PropBright}}
The brightnesses resulting from our parameterized self-gravity wakes are shown in \Fig{}~\ref{dcr_brightness}.  We find for both reflection and transmission that the calculated brightnesses are quite low compared to the brightnesses expected from an unperturbed ring patch lacking self-gravity wakes.  In many cases, this is true even if the unperturbed patch has a lower input mean optical depth than the patch with self-gravity wakes.  

These results lend preliminary quantitative support to a mechanism previously proposed by \citet{Propellers08}, developing an idea first briefly suggested by \citet{Propellers06}, for why propellers appear as bright relative to the background ring.  The argument in our previous paper was that self-gravity wakes tend to lock up ring material into a photometrically inactive state, and that propellers can release this material by locally and temporarily disrupting the wake structure.  Thus, even though a propeller structure contains less material overall \citep{Seiss05,LS09}, it may contain more \textit{photometrically active} material and thus be relative-bright. 

It should be noted that this hypothesis cannot be directly tested at this time by numerical simulations.  To our knowledge, the most detailed simulations carried out to date on propellers are those by \citet{LS09}, but even their results do not fully incorporate the activity of self-gravity wakes, due to the great difficulty in simultaneously accounting for phenomena on very different lengthscales (a few meters for self-gravity wakes, several km for propellers).  The numerical simulations and photometric models in the online supplement of \citet{Sremcevic07} also are insufficient to either confirm or reject this hypothesis due to the severely limited extent of phase space explored.  Investigating small-scale structure requires a large number of particles per unit area, while the large patch size required to investigate large-scale structure drives the total number of particles quite high.  Our hope is that the semi-analytical treatment presented here will provide guidance to the highly computationally-intensive direct models that would be required to quantitatively address this question. 

Other possible hypotheses of course exist, such as the suggestion of \citet{Sremcevic07} that densities within a propeller structure are enhanced by temporarily liberated ring-particle regolith, but it is not within the scope of this paper to study them.  Our hypothesis was very briefly rebutted\fn{in response to informal discussion during 2006, as well as the brief mention by \citet{Propellers06}} in the online supplement of \citet{Sremcevic07}, who rejected the idea primarily because of a general sense that it should lead to ``moonlet wakes''\fn{See \citet{Propellers08} and references therein for more in the context of propellers on the meaning of the term ``moonlet wakes,'' originally due to \citet{Show86}, which has nothing to do with the ``self-gravity wakes'' discussed elsewhere in this paper, despite an unfortunate similarity in the names.} even brighter than the propeller gaps.  However, many realistic simulations of propeller structures lack moonlet wakes \citep{LS09}, and in any case, it may well be that both wakes and gaps are seen as bright in observed propellers.

Our basic suggestion is that the continuum brightness of the ring is characterized by the solid black lines in \Fig{}~\ref{dcr_brightness}, especially in the range of input mean surface densities from 40~to 45~g~cm$^{-2}$ that typify the ``propeller belt'' region \citep{soirings}, and that the brightness of well-developed propeller-shaped features is characterized by the dotted line.  Although the dotted lines in \Fig{}~\ref{dcr_brightness} are above the solid black lines for any given surface density, the comparison must be made with the level of the dotted line at a lower surface density than for the solid black line, since the relative density inside well-developed propeller-shaped features is lower than the background -- e.g., 10\% to 30\% in the models of \citet{LS09}.  Propellers would be relative-bright features if the dotted lines at 10~to 15~g~cm$^{-2}$ have higher relative brightness than the solid black lines at the highest surface densities plotted.  

In keeping with the hypothesis of \citet{Propellers06,Propellers08}, \Fig{}~\ref{dcr_brightness}b shows that all three parameterized optical depth values (colored solid lines) for images in transmission correspond to brightnesses lower than those of unperturbed rings with no wakes (dotted line), even if the latter are considered to have significantly lower optical depths overall.  The low values of the area-weighted average brightness make it quite likely that propellers may be seen as relative-bright features in images of the unlit side of the rings.  For images in reflection (\Fig{}~\ref{dcr_brightness}a), the wakes themselves are quite bright (green symbols), and an average between only the dense wakes and the intermediate optical depth (``$\tau_{gap}$'') would certainly result in wakes that are relative-dark features, but in our results the fact that a large fraction of the area is taken up by effectively empty space pushes the mean brightness (black solid line) downward to a regime in which propellers might be neutral or perhaps relative-bright features, when seen on the lit side of the rings.  

To be sure, we have not shown that this mechanism works in all cases.  Factors that make it more likely to work include 1) higher densities within the propeller gaps, 2) higher background surface density $\sigma$, and 3) lossier coefficient of restitution.  In some scenarios, the propeller gaps might be brightened by this mechanism only enough to blend in with the background, or may remain relative-dark.  It must also be remembered that our analysis is based on several approximations; thus, we cannot say in detail under which conditions this mechanism can or cannot work.  Our purpose is to show that the mechanism is plausible, and to encourage more detailed work.

These conclusions fit well with the observations.  \citet{Propellers08} found that propellers on both the lit and unlit faces of the rings were relative-bright features, and in fact found no relative-dark features in the ``propeller belt'' of the mid-A ring.  Furthermore, \citet{Propellers08} found no significant difference in propeller dimensions between lit-side (reflection) and unlit-side (transmission) images, indicating that the same basic structures were being seen in both geometries.  Yet in morphology the observed propellers strongly resemble the regions of relatively low density seen in simulations, not the relatively dense ``moonlet wakes'' associated with the central moonlet.

Preliminary observations of very large propellers by \citet{DPS08Props} do show, for the first time, some relative-dark regions within propellers.  These ``giant propellers'' occur between the Encke Gap and the Keeler Gap, farther from the center of Saturn than the ``propeller belt'' in the mid-A~ring that contains the smaller propellers observed previously.  But even these observations can be incorporated into the above model simply as a matter of degrees, in effect becoming the exception that proves the rule.  If relative densities inside the giant propellers are exceedingly low, then they will indeed be dark relative to the background in both reflection and transmission (\Fig{}~\ref{dcr_brightness}).  Similarly, if relative densities inside the ``moonlet wakes'' are exceedingly high for the giant propellers, they may become so opaque that they are indeed relative-dark for images in transmission.  However, neither of those circumstances affects our conclusion for the smaller propellers in the ``propeller belt.''\fn{Also, background surface densities are lower beyond the Encke Gap than they are in the ``propeller belt'' \citep{Spilker04}.}  For these, we have found that propellers may very reasonably be found to be relative-bright features, both for images in reflection and for images in transmission, if they locally and temporarily disrupt the structure of self-gravity wakes in addition to decreasing the overall local density within the propeller-shaped structure. 

\section{Summary and Conclusions \label{Conclusions}}
We have developed a semi-analytic method of parameterizing simulations of self-gravity wakes in Saturn's rings, describing their photometric properties by means of only six numbers:  three optical depths and three weighting factors.  These numbers are obtained by fitting a sum of three gaussians to the results of a density-estimation procedure that finds the frequencies of various values of local density within a simulated ring patch.  In order to account for the conversion from dynamical optical depth to photometric optical depth, we use the expression $\tau_{phot}/\tau_{dyn} \simeq 1 + kD$ \citep{SK03}, where $D$ is the volume filling factor (which we estimate); we find the best results by setting the scalar $k=1$.  

Our first surprising result is that rings dominated by self-gravity wakes appear to be mostly empty space, with more than half of their area taken up by local optical depths around~0.01.  Such regions will be photometrically inactive for all viewing geometries.  While this result might be affected by our use of a monodisperse size distribution, we suggest that it might be robust due to the lower coefficient of restitution that we used in our $N$-body simulations, following \citet{Porco08}. 

If our models are in fact robust, then the bimodal density distribution assumed in the interpretation of occultation data by previous investigators should be replaced by a \textit{tri}modal distribution.  The practical results of this turn out to be minimal, as occultations can only distinguish between bimodal and trimodal models at very low opening angle.  Existing analysis of occultation data should be interpreted with $\tau_{gap}$ representing the area-weighted average optical depth within the gaps (or inter-wake regions), keeping in mind the possibility that there may be strong variations in optical depth within those inter-wake regions. 

Applying our parameterization of self-gravity wakes lends preliminary quantitative support to the hypothesis of \citet{Propellers08} that ``propellers'' observed in the mid-A ring are bright because of a local and temporary disruption of self-gravity wakes.  Even though the overall local density is lower within the propeller-shaped structure surrounding an embedded central moonlet, disruption of the wakes would flood these same regions with more photometrically active material, raising their apparent brightnesses in agreement with observations.  We find that this mechanism can plausibly work for a wide range of input parameters.

We hope that this hypothesis will eventually be directly tested by detailed numerical simulations.  However, this is presently very difficult due to the high expense in terms of computational resources necessary to account for the small lengthscales appropriate for self-gravity wakes (a few meters) while simultaneously accounting for the very large lengthscales appropriate for propellers (several km).  The former essentially requires a large number of particles per unit area, while the latter requires a large patch size.  As computational resources increase to the point that such simulations become feasible, it will become possible to determine whether propellers really are characterized by a local and temporary disruption of self-gravity wakes, as well as whether the photometric properties of propellers can be explained using the hypothesis we have outlined.  In the meantime, the results of our semi-analytic method give us reason to hope that the solution to the problem indeed lies in this direction. 

\acknowledgements{We thank Tom~Loredo, Hyunsook~Lee, Mark~Lewis, and Phil~Nicholson for helpful comments, and an anonymous reviewer for helping us to improve the manuscript.  M.S.T. acknowledges funding from NASA's Cassini Data Analysis Program (NNX08AQ72G).  R.P.P. acknowledges support from the NASA Earth and Space Science Fellowship (NESSF) Program.  D.C.R. acknowledges funding from NASA grant NNX08AM39G, issued through the Office of Space Science.  J.W.W. and C.C.P. and D.C.R. acknowledge funding from NASA's Cassini Data Analysis Program (NNX08AP84G).}

\bibliographystyle{apalike}
\bibliography{bibliography}

\end{document}